\documentclass[aps,prl,reprint,showpacs,floatfix,a4paper,superscriptaddress]{revtex4-1}
\usepackage{graphicx}
\usepackage{amsmath}
\usepackage{amssymb}
\usepackage{amsfonts}
\usepackage{color}
\usepackage{relsize}
\usepackage{mathrsfs}
\usepackage{txfonts}
\usepackage[T1]{fontenc}
\usepackage{microtype}
\usepackage[pdfpagemode=UseNone,colorlinks=true,linkcolor=blue,citecolor=blue]{hyperref}
\usepackage{siunitx}

\renewcommand\Re{\operatorname{Re}}
\renewcommand\Im{\operatorname{Im}}

\newcommand{\ii}{\mathrm{i}}

\newcommand{\ee}{\mathrm{e}}

\newcommand{\rmt}{\mathrm{t}}
\newcommand{\rmr}{\mathrm{r}}

\newcommand{\frake}{\mathfrak{e}}

\newcommand{\calP}{{\mathcal P}}
\newcommand{\calC}{{\mathcal C}}
\newcommand{\calR}{{\mathcal R} }

\newcommand{\calS}{{\mathcal S}}
\newcommand{\calF}{{\mathcal F}}

\newcommand{\calL}{{\mathcal L}}
\newcommand{\calD}{{\mathcal D}}

\newcommand{\calI}{{\mathcal I}}

\renewcommand{\Re}{\mathfrak{Re}}
\renewcommand{\Im}{\mathfrak{Im}}

\newcommand{\tmf}{{\mathfrak t}}
\newcommand{\rmf}{{\mathfrak r}}

\begin{document}

\title{Two--membrane etalon}

\author{F.~Marzioni}
\affiliation{School of Science and Technology, Physics Division, University of Camerino, I-62032 Camerino (MC), Italy}
\affiliation{INFN, Sezione di Perugia, via A. Pascoli, I-06123 Perugia, Italy}
\affiliation{Department of Physics, University of Naples “Federico II”, I-80126 Napoli, Italy}
\author{R.~Natali}
\affiliation{School of Science and Technology, Physics Division, University of Camerino, I-62032 Camerino (MC), Italy}
\affiliation{INFN, Sezione di Perugia, via A. Pascoli, I-06123 Perugia, Italy}
\author{N.~Malossi}
\affiliation{School of Science and Technology, Physics Division, University of Camerino, I-62032 Camerino (MC), Italy}
\affiliation{INFN, Sezione di Perugia, via A. Pascoli, I-06123 Perugia, Italy}
\author{D.~Vitali}
\affiliation{School of Science and Technology, Physics Division, University of Camerino, I-62032 Camerino (MC), Italy}
\affiliation{INFN, Sezione di Perugia, via A. Pascoli, I-06123 Perugia, Italy}
\affiliation{CNR-INO, Largo Enrico Fermi 6, I-50125 Firenze, Italy}
\author{G.~Di~Giuseppe}
\affiliation{School of Science and Technology, Physics Division, University of Camerino, I-62032 Camerino (MC), Italy}
\affiliation{INFN, Sezione di Perugia, via A. Pascoli, I-06123 Perugia, Italy}
\author{P.~Piergentili} \email[Corresponding author:] {paolo.piergentili@unicam.it}
\affiliation{School of Science and Technology, Physics Division, University of Camerino, I-62032 Camerino (MC), Italy}
\affiliation{INFN, Sezione di Perugia, via A. Pascoli, I-06123 Perugia, Italy}


\begin{abstract}
Optomechanics with semi--transparent membrane multi--oscillators in a high--finesse cavity is an established solution for designing the dispersive interaction, and reaching many achievements, such as the study of non--linear dynamics, heat transfer, and so on.
The multi--oscillators are dielectric slabs, usually with low reflectivity, constituting an etalon.
Here we propose the theoretical and experimental investigation of a low-finesse optical cavity formed by two (nominally identical) parallel semi–transparent membranes.
The experiment consists in laser-driving the cavity, detecting interferometrically the field reflected by the etalon, and measuring the intensity of the transmitted field, for different distances within the two membranes.
A complete characterization of the membrane sandwich is provided.
On the other hand, we developed an analytical model to describe the fields, which reproduces all the experimental results, under specific approximations.
As expected, the model reproduces the known results when high–reflectivity, and/or fixed mirrors conditions are restored.
This work paves the way for a complete and analytical model to describe multi-oscillators ``membrane-in-the-middle'' optomechanics.
\end{abstract}


\maketitle

\section{Introduction}
In the second half of the XX century, the idea of building up optical eigenmodes within two reflective surfaces paved the way to the realization of lasers \cite{Kogelnik:66,Siegman:1999}.
Since then, optical resonators, or cavities, acquired a fundamental role in many areas of Physics \cite{YE20031}, and the problem of quantizing the cavity field has been addressed \cite{PhysRevA.12.148,Kim:1994vs,Aiello:2000wn,Viviescas:2003}.
Optomechanics, where optical resonators are coupled to mechanical oscillators \cite{Aspelmeyer:2014vr}, allows the exploration of quantum mechanics \cite{Teufel:2011vw,Verhagen:2012aa,Brubaker:2022aa}, and the quantum control of mechanical motion \cite{Rossi:2018ve}. 
The detection of weak forces and displacements has been realized in optomechanical systems \cite{Mason:2019aa}, which are also one the most promising platforms for the implementation of quantum technologies \cite{Barzanjeh:2022aa}.
The ``membrane in the middle'' (MIM) configuration \cite{Thompson:2008aa,Jayich_2008,PhysRevLett.103.207204}, which consists of a semi--transparent dielectric slab mounted inside an optical cavity, is a rife solution to achieve the optomechanical dispersive interaction.
The variety of applications of this kind of setup grows up if the possibility of coupling more than one mechanical oscillator to the same optical field is considered \cite{PhysRevA.78.041801,Gartner:2018aa,PhysRevA.99.023851}.
To give a few examples, it has been proven that the single--photon optomechanical coupling, which quantifies the strength of the interaction, can be enhanced if two membranes are coupled to the same optical field \cite{Piergentili:2018aa,PhysRevX.15.011014}.
Heat transport \cite{Yang:2020aa}, non--linear dynamics \cite{Piergentili:2021,PhysRevApplied.15.034012,Piergentili:2022aa}, coherent noise cancellation \cite{10.3389/fphy.2023.1222056}, exceptional points \cite{Xu:2016aa}, have been observed in multi--oscillator systems.
In these ensemble of systems the multi--oscillator array forms inner cavities, located within an high--finesse optical resonator.
The measurable fields depend heavily on the configuration of the inner cavities, and the knowledge of their transfer function is useful for a complete analysis of multi--oscillator MIM systems.

Here we propose the theoretical and experimental investigation of an optical cavity formed by two (nominally identical) parallel semi--transparent membranes.
In this scenario the ``good--cavity'' approximation does not hold, due to the low reflectivity of the mirrors, and the oscillations of the membranes introduce dynamical boundary conditions for the cavity field. 
We present a model which overtakes these difficulties, it is consistent with the experiment, and it can be considered a generalization, since it leads to the known results when high--reflectivity, and/or fixed mirrors conditions are restored.

In the first section we develop the theoretical model.
We approach the problem analytically, writing down the equations for the intra--cavity, reflected, and transmitted fields.
The described method leads to an expression for the field spectral density around the mechanical modes of the membranes.
The second section provides a description of the experimental apparatus, we show the measured spectra, and we compare them to the numerical simulations. 
The main results of the work are discussed.
In the third section we provide a complete characterization of the two--membrane etalon.
\section{Theory}
We consider the case of two different movable dielectric membranes forming a Fabry--P\'erot cavity of length $L$, which is driven by an external laser~\cite{Piergentili:2018aa}. The membranes can be modelled as dielectric slabs of thickness $L_{{\rm m},j}$ and index of refraction $n_j$ (where the index $j=1,2$ distinguish the parameters of the two membranes), such that their reflection and transmission coefficient can be expressed as
\begin{eqnarray}
    \rmf_j = \frac{(n_j^2-1)\sin(kn_jL_{{\rm m},j})}
    				{(n_j^2+1)\sin(kn_jL_{{\rm m},j}) +  2\ii\, n_j\cos(kn_jL_{{\rm m},j})} 	
	\label{eq:rm}\\
    \tmf_j = \frac{2n_j}
    				{(n_j^2+1)\sin(kn_jL_{{\rm m},j}) +  2\ii\, n_j\cos(kn_jL_{{\rm m},j})}		
	\label{eq:tm}		
\end{eqnarray}
where $k =\omega/c = 2\pi/\lambda$ is the wavenumber of the electric field, and $\lambda$ is its wavelength.
In the case of negligible optical absorption of the membranes, i.e. for real $n_j$, we can rewrite $\rmf_j$ with $j=1,2$ in terms of the intensity reflectivity ${\mathfrak R}_j$ as $\rmf_j=\sqrt{{\mathfrak R}_j}e^{\ii\, \varphi_j}$.
\begin{figure}[h!]
\begin{center}
   {\includegraphics[width=.315\textwidth]{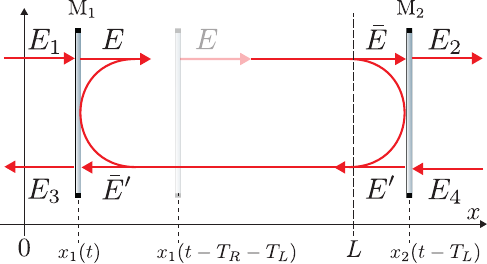}}
 \caption{Two--membrane cavity fields. The two dielectric slabs, indicated as M$_1$ and M$_2$, are positioned at $x_1$ and $x_2$, respectively. They form a Fabry-P\'{e}rot cavity, having input fields $E_1$ and $E_4$, and output fields $E_2$ and $E_3$. Inside the etalon, we consider $E$, which propagates to the right, and $E^\prime$ to the left. We indicate a propagation through the cavity with the bar symbol. The zero for the phases of the fields is taken at $x=0$. The motion of the membranes introduces dynamical boundary conditions for the fields. The field at time $t$ brings the information on the slabs' position at the instant of their interaction, which is retarded by the period the light needs to travel across the cavity: $T_L$ if the fields propagates to the left, $T_R$ to the right.}
\label{fig:Figure_TwoMemCavity}
\end{center}
\end{figure}

According to Fig.~\ref{fig:Figure_TwoMemCavity}, we assume time dependent input/output fields, with bandwidth much smaller than the carrier frequency $\omega_L$ (quasi--monochromatic approximation):
 \begin{equation}
 	\mathcal{E}_l(x,t) = E_l(t)\,\ee^{\ii\omega_L t + \ii \phi(x)}\,,\quad l=1,2,3,4\,.
 \end{equation}
We choose the zero of the phases, acquired by the fields traveling a distance $x$, at the input of the cavity positioned at $x=0$. According to $\phi_L(x) = -\phi_R(x) =  kx = \omega_L x/c$, where the indexes $(R,L)$ indicate field propagating forward (right) or backward (left), the slowly--varying amplitude fields satisfy, on the first mirror, the boundary conditions 
\begin{align}
	&E(t)\,\ee^{-\ii k x_1(t)} = \tmf_1\, E_{1}(t) \,\ee^{-\ii kx_1(t)}
						       + \rmf_1\,\bar E^\prime(t)\,\ee^{\ii k x_1(t)}     \label{eq:M11}
		\\
	&E_{3}(t)\,\ee^{\ii k x_1(t)}  =  -\rmf_1^\ast\, E_{1}(t) \,\ee^{-\ii k x_1(t)}
								  + \tmf_1^\ast\,\bar E^\prime(t)\,\ee^{\ii k x_1(t)}   \label{eq:M12}
	\,.
\end{align}
On the second mirror
\begin{align}
	E_{2}(t)\,\ee^{-\ii k x_2(t)}  &=  \tmf_2\,\bar E(t)\,\ee^{-\ii k x_2(t)} + \rmf_2	\, E_4(t)\,\ee^{\ii k x_2(t)} \label{eq:M21}
		\\
	E^{\prime}(t)\,\ee^{\ii k x_2(t)} &= -\rmf_2^\ast \,\bar E(t)\,\ee^{-\ii k x_2(t)}	+ \tmf_2^\ast \, E_4(t)\,\ee^{\ii k x_2(t)} \label{eq:M22}
	\,.
\end{align}
The slowly varying amplitudes at the mirror positions are related each other by~\cite{Saleh:2007sw}
\begin{align}
	\bar E(t)&= E(t-T_R) \label{eq:Eb}\\
	\bar E^{\prime}(t)&= E^\prime(t-T_L)	\label{eq:Ebp}\,,
\end{align}
with $T_R= [x_2(t-T_L)-x_1(t- T_L -T_R)]/c$, and $T_L =[x_2(t - T_L)-x_1(t )]/c$, where $c$ is the speed of light.

We assume in the following $E_4(t) = 0$, for which
\begin{align}\label{eq:Ep}
	E^\prime (t) &= -\rmf_2^\ast \,\ee^{-2\ii k x_2(t)} \,\bar E(t)
	\,,
\end{align}
and evaluate the cavity field  $E(t)$ in Eq.~\eqref{eq:M11} by using relations~\eqref{eq:Eb}-\eqref{eq:Ebp}, and Eq.~\eqref{eq:Ep}
\begin{align}\label{eq:E2}
	E(t) = \tmf_1 E_1(t) -\rmf_1\rmf_2^\ast &E(t - T_L -T_R) \\
	&\times\ee^{-\ii k [2x_2(t-T_L)-x_1(t) - x_1(t -T_L -T_R)]} \nonumber
	\,.
\end{align}

The instantaneous round--trip time $ T_L + T_R$ changes due to the membrane displacements, $x_1(t)$ and $x_2(t)$, according to $2[L + \delta l(t)]/c$ where $L$ is the stationary cavity length, and $\delta l(t) = 2\delta x_2(t-T_L)-\delta x_1(t) - \delta x_1(t -T_L -T_R)$ is the time--dependent cavity length variation with zero mean--value. 
In the following, $L=L_0+\delta L$, where $\delta L\ll L_0$ is a stationary cavity--mismatch length with respect to $L_0$, the cavity length for the resonance at the cavity frequency $\omega_c$. 
For typical mechanical frequencies in the range $\omega_m/2\pi = \SI{100}{\kilo\hertz}-\SI{10}{\mega\hertz}$ corresponding to a period time $T_m = \SI{100}{\pico\s}$ -- \SI{10}{\micro\s} much longer than round--trip time $\tau < \SI{7}{\pico\s}$ for cavity length $L_0 <\SI{1}{m}$, we can assume that the intracavity electric field probes both the membranes, in a round-trip, instantaneously, and as a result  $T_R \simeq T_L \simeq \left[x_2(t) - x_1(t)\right]/c = \tau/2 + \delta l(t)/c$, with $\tau = 2(L_0 + \delta L)/c=1/{\rm FSR}$ and $\delta l(t) = 2[\delta x_2(t)-\delta x_1(t)]$. Moreover, assuming $|\delta x_j| \ll L_0+\delta L$, the equation for the intracavity slowly--varying amplitude can be cast as
\begin{align}\label{eq:E3}
	E(t) = \tmf_1 E_1(t) + \rmf_1 \bar \rmf_2^\ast E(t - \tau)\,\ee^{-\ii\Phi(t)} 
	\,,
\end{align}
with $\bar\rmf_2 = -\rmf_2 \,\ee^{\ii \omega_L \tau }$, $\Phi(t) = 2k[ \delta x_2(t)- \delta x_1(t)]$. Such result is equivalent to the dynamics derived in Ref.~\cite{Sacher:2008aa} for a microring modulator in the presence of loss and refraction index modulations but with a constant input field.

From the cavity field  $E(t)$ we can derive the transmitted and reflected fields, that is $E_t(t)\equiv E_2(t)$ and $E_r(t) \equiv E_3(t)$, respectively. The former is evaluated inserting Eq.~\eqref{eq:Eb} in Eq.~\eqref{eq:M21} for $E_4 =0$
\begin{align}\label{eq:Et}
	E_t(t) = \rmt_2E(t - \tau/2)
	\,,
\end{align}
and the reflected field is provided by the expression
\begin{align}\label{eq:Er2}
	E_r(t) = - \rmr_1^\ast E_1(t) \,\ee^{-2\ii k\delta x_1(t)} + \rmt_1^\ast \bar\rmr_2^\ast E(t - \tau) \,
				\ee^{-2\ii k\delta x_2(t)} 
	\,.
\end{align}
We note that this expression provides the results for a single membrane when setting one of the reflectivity equal to zero and transmittivity to one. In general, the first membrane modulates directly the input field, while the second membrane interacts with the electric field mediated by the cavity response.

In time domain, Eq.~\eqref{eq:E3} can be express as a Fredholm linear integral equation of the second kind~\cite{Crosignani:1986aa,Sacher:2008aa}
\begin{align}\label{eq:E4}
	E(t) = \mu \int_{-\infty}^{+\infty} K(t,t^\prime)\, E(t^\prime)\,dt^\prime + F(t) \,
	\,,
\end{align}
with $F(t) = \tmf_1 E_1(t)$, $\mu  = \rmf_1\bar\rmf_2^\ast \,$, and $K(t,t^\prime) = \delta[t^\prime - (t - \tau)]\,\exp[-\ii \Phi(t^\prime)]$, which possesses a Neumann series solution~\cite{Krasnov1983:aa} (the series converges for $\mu<B^{-1}$, with $B^2 = \int\int K^2(t, t^\prime)dt dt^\prime$)
\begin{align}\label{eq:E5}
	E(t) = F(t) + \sum_{n =1}^{\infty} \mu^n \int_{-\infty}^{+\infty}K_n(t,t^\prime) \,F(t^\prime)\,dt^\prime
	\,,
\end{align}
where $K_1(t,t^\prime) = K(t,t^\prime)$, and
\begin{align}\label{eq:E6}
	K_n(t,t^\prime) = \int_{-\infty}^{+\infty}K(t,s)\, K_{n-1}(s,t^\prime) \,ds \,
	\,.
\end{align}
For the kernel of Eq.~\eqref{eq:E4} we have $K_n(t,t^\prime) = \delta[t^\prime-(t-n\tau)]\prod_{m=0}^{n-1}\exp[-\ii\Phi(t-m\tau)]$,  and  the cavity field $E(t)$ is given by
\begin{align}\label{eq:E7}
	E(t) =\tmf_1 \sum_{n =0}^{\infty} 
		\left(\rmf_1\bar\rmf_2^\ast\right)^n
	\, E_1(t - n\tau)\,\prod_{m=0}^{n-1}\,\ee^{-\ii\Phi(t-m\tau)} \,
	\,.
\end{align}

From the cavity field  $E(t)$ we can derive the transmitted and reflected fields, that is $E_2(t)$ and $E_3(t)$, respectively,
\begin{align}\label{eq:Et}
	E_t(t)\equiv E_2(t) =\tmf_1 \tmf_2\sum_{n =0}^{\infty}
		(\rmf_1\bar\rmf_2^\ast)^n
		E_1&(t -\tau/2 - n\tau)
			\nonumber\\
		\times&\prod_{m=0}^{n-1}\,\ee^{-\ii\Phi(t-\tau/2 - m\tau)}		
	\,,
\end{align}
and the reflected field is provided by the expression
\begin{align}\label{eq:Er}
	E_r(t) \equiv &E_3(t) = -\rmf_1^\ast E_1(t) \,\ee^{-2\ii k\,\delta x_1(t)}
			 \\
	&+\frac{|\tmf_1|^2}{\rmf_1}\,\ee^{-2\ii k\,\delta x_2(t)} \sum_{n =1}^{\infty} 
	\left(\rmf_1\bar\rmf_2^\ast\right)^n\, 
	E_1(t  - n\tau)\,\prod_{m=0}^{n-1}\,
	\ee^{-\ii\Phi(t - m\tau)} \nonumber
	\,,
\end{align}
that is, the field consists of an instantaneous response, and a summation of memory terms delayed by the round-trip time and weighted by $\rmf_1\bar\rmf_2^\ast=-\rmf_1\rmf_2^\ast\,\ee^{-\ii \omega_L \tau }$. This general expression reproduces Eq.~(3) in Ref.~\cite{Tsubono:1991aa} for a monochromatic input field $E_1 = a$, by assuming $x_1 = 0$, and $\rmf_j = \ii r_j$. It also reproduces Eq.~(33) in Ref.~\cite{Raymer:2013aa} for static mirrors, by assuming $E_1= -A(t)$, $x_2-x_1 = \tau c/2$, $\rmf_2 = 1$, $\rmf_1 = \rmf_1^\ast = -\rho$.

\section{Signal detection}
According to the experimental setup of Fig.~\ref{fig:Figure_Experiment}, a laser beam is split to obtain the low--intensity input field $E_{1}(t)= \sqrt{\calP_{in}}\,f(t)$, and a strong field $E_{lo}(t)= \sqrt{\calP_{lo}}\,f(t) \exp(\ii\phi_{l})$, local oscillator, phase shifted by $\phi_{l}$ with respect to the field $E_{1}(t)$. The homodyne detection is obtained by interfering on an ideal $50/50$ beam--splitter the reflected field, given by $E_{r}(t)$, with the local oscillator.  
The two output fields are given by $E_{\pm}(t) = [ E_{r}(t) \pm  E_{lo}(t)\,\ee^{\ii\phi_l}]/\sqrt{2}$. 
The differential detection photocurrent, assuming a balanced interferometer, that is the path difference is much less than the coherence length of the laser, is given by:
\begin{align}
	\calI(t) = 2\sqrt{\calP_{lo}\calP_{in}}\cdot
		\frake(t)
		\,,
\end{align}
with a time--dependent contribution
\begin{align}\label{eq:Efrake}
	\frake(t)   	=\Re\left\{ f^\ast(t)\,\ee^{-\ii\phi_l}\,\frac{E_r(t)}{\sqrt{\calP_{in}}}\right\} 
			\,.
\end{align}
It provides a voltage signal
\begin{align}
	\tilde V_H(t) = g_T\,S \cdot2\sqrt{\calP_{lo}\calP_{in}}\cdot
		\frake(t)
		\,,
\end{align}
where $S\,[\si{\ampere\per\watt}]$ is the responsivity of the photodiodes, $g_T\,[\si{\volt\per\ampere}]$ the gain of a transimpedance amplifier with bandwidth $\Omega_T\ll \omega_m$.

The physical information is gained by detecting the noise spectral density of the homodyne signal, $\calS_W( \omega)$, that is the Fourier transform of the correlation function $G_{VV}(\theta)= \langle \tilde V_H^\ast(t + \theta)\tilde V_H(t) \rangle$:
\begin{align}
	\calS_{W}(\omega) &= \int d\theta\,G_{VV}(\theta)\,\ee^{\ii \omega \theta} \\
				      &= (g_TS)^2 \cdot4\calP_{lo}\calP_{in}\cdot
		\int d\theta\, \ee^{\ii\omega \theta}
		\,\langle\frake(t+\theta)\frake(t)\rangle_t
		\,.
\end{align}
It can be cast as ($s=\ii\omega$)
\begin{align}
	\calS_W( s) 
	= (g_TS)^2 \cdot4\calP_{lo}\calP_{in}\cdot
		\big|\tilde\frake(s)\big|^2		
		\,,
\end{align}
with $\tilde\frake(s)$ the Laplace transform of the expression in Eq.~\eqref{eq:Efrake}, which requires the knowledge of the transform for the fields.

The Laplace transform of the cavity field in Eq.~\eqref{eq:E3} yields 
\begin{align}\label{eq:Ecvs}
	\tilde E(s) = \rmt_1 \tilde E_{1}(s)  
		+ \rmr_1\bar\rmr_2^\ast\,\calL\Big\{E(t-\tau)\ee^{-\ii k\delta l(t)}\Big\}
		\,.
\end{align}
Assuming displacements with amplitude $\delta \tilde x_j(\omega_{mj})$ at frequency $\omega_{mj}$, that is $\delta x_j (t) = \delta x_j(\omega_{mj})\sin(\omega_{mj} t )$, the solution of Eq.~\eqref{eq:Ecvs} can be found considering the Jacobi-Anger expansion  of the exponential in terms of the parameters $\xi_j = 2\omega_L \delta x(s_{m}) /c$, for which $\exp{[-\ii\xi\sin(\omega_m t)]} = \sum_m(-1)^m J_m(\xi)\exp{(ms_m t)}$ with $s_m = \ii\omega_m$, and $J_m $ the first kind Bessel function of order $m$ 
\begin{align}\label{eq:Ecvs2}
	\tilde E(s) = & \rmt_1 \tilde E_{1}(s)  + \rmr_1\bar\rmr_2^\ast\,\ee^{-s\tau}
			\\
		\times&
		\sum_{m,n}(-1)^n J_n(\xi_1)J_m(\xi_2)
		\tilde E(s-ns_{n1} -ms_{m2})\,\ee^{(ns_{n1}+ms_{m2})\tau}		\nonumber\,,
\end{align}
which can be cast in the form 
\begin{align}\label{eq:Ecvs3}
	\tilde E(s)\,&\calD(s)  =  \rmt_1 \tilde E_{1}(s)   + \rmr_1\bar\rmr_2^\ast\,\ee^{ -s\tau}
			\\
		\times&
		\sum_{m,n\neq0}(-1)^m J_n(\xi_1)J_m(\xi_2)
		\tilde E(s-ns_{n1} -ms_{m2})\,\ee^{(ns_{n1}+ms_{m2})\tau}	\nonumber\,,
\end{align}
where 
\begin{align}\label{eq:Ds}
	\calD(s) = \Big[1 - \rmr_1\bar\rmr_2^\ast\,\ee^{-s\tau} J_0(\xi_1)J_0(\xi_2)\Big]
		\,.
\end{align}

The same approach can be followed for the reflected field in Eq.~\eqref{eq:Er2}
\begin{align}
	E_{r}(t) = -\rmr_1^\ast\,\ee^{ - \ii\xi_1(t)}E_{1}(t) + \rmt^\ast_1 \bar\rmr_2^\ast\,\ee^{- \ii\xi_2(t)}E(t-2T)	
	\,,
\end{align}
where the Laplace transform is given by
\begin{align}\label{eq:Erfs}
	\tilde E_{r}&(s) = -\rmr_1^\ast\,\sum_{m}(-1)^mJ_m(\xi_{1})\,\tilde E_{in}(s - ms_{m1})\,\ee^{ms_{m1}T}
				\nonumber\\
		&+ \rmt_1^\ast \bar\rmr_2^\ast\,\ee^{ -s2T} \sum_m(-1)^m J_m(\xi_{2})\tilde E(s -ms_{m2})\,\ee^{ms_{m2}T}
		\,,
\end{align}
and we have introduced the input field $\tilde E_{in}(s)\equiv\tilde E_{1}(s)$.

A perturbative expansion can be sought for $\xi \ll 1$ considering $\tilde E(s) = \sum_{k,p}\xi_1^k\xi_2^p\tilde E^{(k+p)}(s)$, and $J_m(\xi) \sim (\xi/2)^m/m!$ for $m$ integer. The lower orders expansion of the cavity--field is
\begin{align}\label{eq:EcvsApp1}
	\tilde E^{(0)}(s) = \calC^{(0)}(s)\, \tilde E_{in}(s) 
		\qquad\qquad 
	 \calC^{(0)}(s) & =\frac{t_1}{\calD(s)}
			\,,
\end{align}
and
\begin{align}\label{eq:EcvsApp2}
	\xi_j\,\tilde E^{(1)}(s) = \xi_j \, (-1)^j\, \Bigg[  &\calC^{(1)}_j(s,+ s_{mj}) \tilde E_{in}(s+s_{mj})
			\\ 
		 &
		 - \calC^{(1)}_j(s, - s_{mj}) \tilde E_{in}(s -s_{mj})\Bigg]
			\nonumber\,,
\end{align}
where 
\begin{align}
	 \calC^{(1)}_j(s,\pm s_{mj}) &=\frac{t_1}{2} \frac{[1-\calD(s)]}{\calD(s)}\, \frac{\ee^{\mp s_{mj}T}}{\calD(s\pm s_{mj})}
\end{align}

We note that this result reproduces the solution obtained for a nearly resonant high--finesse cavity~\cite{Schliesser:2008hc}, and assuming $\omega_m \ll 2\pi{\rm FSR}$~\cite{Reinhardt:2017ab}, with the approximated expression
\begin{align}
	\calD(s) = \Big[\kappa + \ii\big(\Delta + \omega_L\delta L/c\big) +  \ii\omega\Big]/{\rm FSR}
	\,,
\end{align}
where $\kappa = \kappa_1 + \kappa_2$, and $2\kappa_j = |t_j|^2/{\rm FSR}$. We also note that usually the intracavity field is normalized such that to have a energy density $\tilde a(s) = \tilde E(s)/\sqrt{\rm FSR}$.

By using Eq.~\eqref{eq:EcvsApp1}, the lower order expansion term for the reflected field is
\begin{align}\label{eq:ErfsApp}
	\tilde E^{(0)}_{r}(s) =\tilde\calR^{(0)}(s,\xi)\, \tilde E_{in}(s)
		\,,
\end{align}
with 
\begin{align}\label{eq:R0}
	\tilde\calR^{(0)}(s)
	=  -\rmr^\ast_1 +  \bar\rmr^\ast_2 \,\frac{ |t_1|^2\,\ee^{-s2T}}{\calD(s)}
	\,,
\end{align}

The first order terms, providing Eq.~\eqref{eq:EcvsApp2}, become
\begin{align}
	\xi_j\,\tilde E_r^{(1)}(s) = \xi_j \, (-1)^j\, \Bigg[  &\calR^{(1)}_j(s,+ s_{mj}) \tilde E_{in}(s+s_{mj})
			\\ 
		 &
		 - \calR^{(1)}_j(s, - s_{mj}) \tilde E_{in}(s -s_{mj})\Bigg]
			\nonumber\,,
\end{align}
where 
\begin{align}
	 \calR^{(1)}_j(s,\pm s_{mj}) =
	 \frac{\rmr_1^\ast}{4}\big[1&-(-1)^j\big]
	 	 \\
	 	&+\frac{\bar\rmr_2^\ast}{2} \frac{[1-\calD(s)]}{\calD(s)}\, \frac{|t_1|^2\,\ee^{-2(s \pm s_{mj})T}}{\calD(s\pm s_{mj})}
		\nonumber\,.
\end{align}
For $s,s_m$ much less than the cavity bandwidth and ${\rm FSR}$, that is \emph{bad--cavity regime}, then 
\begin{align}\label{eq:Ers_App}
	\tilde E_{r}(s) \sim 
	 &\calR^{(0)}(0)\,\tilde E_{in}(s)
	 		\\
	 -& \tilde\calR_1^{(1)}(0)\,\xi_1(s_{m1})\left[
		  \tilde E_{in}(s+s_{m1}) -\tilde E_{in}(s-s_{m1})
		  \right]
	 		\nonumber\\
	 +& \tilde\calR_2^{(1)}(0)\,\xi_2(s_{m2})\left[
		  \tilde E_{in}(s+s_{m2}) - \tilde E_{in}(s-s_{m2})
		  \right]
		  \nonumber\,.
\end{align}

The Laplace transform of expression~\eqref{eq:Efrake}, considering~\eqref{eq:Ers_App}, and assuming $E_{in}(t) = \sqrt{\calP_{in}}\,f(t)$
can be cast as 
\begin{align}
	\tilde \frake(s) = &\frac{\sqrt{\calP_{in}}\,\ee^{-\ii\phi_l}}{2}
		\Bigg\{\calR^{(0)}(0)\,\tilde \calI_{in}(s)
	 		\\
	 -& \xi_1(s_{m1})\,\eta_1\,\tilde\calR_1^{(1)}(0)\left[
		 \tilde \calI_{in}(s - s_{m1}) -\tilde \calI_{in}(s +s_{m1})
		  \right]
	 		\nonumber\\
	 +& \xi_2(s_{m2})\,\eta_2\,\tilde\calR_2^{(1)}(0)\left[
		  \tilde \calI_{in}(s-s_{m2}) - \tilde \calI_{in}(s + s_{m2})
		  \right]
		\Bigg\} + c.c.  
		\nonumber\,,
\end{align}
where $ \calI_{in}(s) = \int ds^\prime\,[\tilde f(s-s^\prime)]^\ast \tilde f(s^\prime)$ is the Laplace transform of the function $|f(t)|^2$, which represents the normalized input intensity, and $ \eta_j = \int d\vec\rho\,  |g(\vec \rho)|^2\,U^{(pq)}_j(\vec \rho)$ is the overlap between the transverse intensity profile of the beam  $|g(\vec\rho)|^2$ and the eigenfunction $U^{(pq)}_j(\vec \rho)$ of the mode $(pq)$ for the $j-$membrane.

For a single--mode, stable laser beam, the Laplace transform of the intensity represents a delta function, that is the homodyne amplitude spectrum has a contribution at low--frequency (DC), and AC tones at $\pm\omega_{mj}$, providing
\begin{align}
	\tilde \frake(\omega) &= \sqrt{\calP_{in}}
		\Bigg \{ \Re\Big[ \calR^{(0)}(0)\,\ee^{-\ii\phi_l}\Big]\,\delta(\omega)
	 		\\
	 -&\xi_1(\omega_{m1})\, \eta_1\,\Re\Big[\calR_1^{(1)}(0)\,\ee^{-\ii\phi_l}\Big]\,
	 	\Big[\delta(\omega - \omega_{m1}) - \delta(\omega + \omega_{m1}) \Big]
	 		\nonumber\\
	 +&\xi_2(\omega_{m2})\, \eta_2\,\Re\Big[\calR_2^{(1)}(0)\,\ee^{-\ii\phi_l}\Big]\,
	 	\Big[\delta(\omega - \omega_{m2}) - \delta(\omega + \omega_{m2}) \Big]
		\Bigg\}
		\nonumber\,,
\end{align}
The mechanical amplitude spectra for the membranes are 
\begin{align}
	\xi_j(\omega) = \frac{2\omega_L}{c}\frac{\tilde F_j(\omega)}{m_{eff}}\frac{1}{\omega_{mj}^2 - \omega^2 + \ii \gamma_{mj}\omega}= \frac{2\omega_L}{c}\chi_{m,j}(\omega)\tilde F_j(\omega)
	\,,
\label{eq:displ_spectra}
\end{align}
where $\chi_{m,j}(\omega)$ is the mechanical susceptibility of the $j-$mode. 
$\tilde F_j(\omega)$ can be considered either uncorrelated, for example in case of thermal noise, or correlated, for example by applying a common mechanical perturbation.
In general then, the single--sided voltage noise spectrum, which is optimal for $\phi_l = \pm \pi/2$, around the mechanical frequencies is 
 \begin{align}
	\calS_W( \omega)  
	= &\left(\frac{4g_TS\omega_L}{c}\sqrt{2\calP_{lo}\calP_{in}}\right)^2
		\bigg\{
		\left[\eta_1\Im\left\{\calR_1^{(1)}\right\}\right]^2\calS_{x_1x_1}(\omega)
			\nonumber\\
		&\hspace{3cm}+ \left[\eta_2\,\Im\left\{\calR_2^{(1)}\right\}\right]^2\calS_{x_2x_2}(\omega)
			\nonumber\\
		&\hspace{1.25cm}-\eta_1\eta_2\Im\left\{\calR_1^{(1)}\right\}\Im\left\{\calR_2^{(1)}\right\}
		\Re\Big\{\calS_{x_1x_2}(\omega)\Big\}		
		\bigg\}
		\,,
\label{eq:Sxx}
\end{align}
where  $\calS_{x_jx_j}(\omega)=|\chi_{m,j}(\omega)|^2$ is the displacement noise spectrum of the $j-$membrane, $\calS_{x_1x_2}(\omega)=\chi_{m,1}^\ast(\omega)\chi_{m,2}(\omega)$ the cross-spectral density noise between the displacement of the two membranes.
\section{Experiment}
\begin{figure}[h!]
\begin{center}
   {\includegraphics[width=0.45\textwidth]{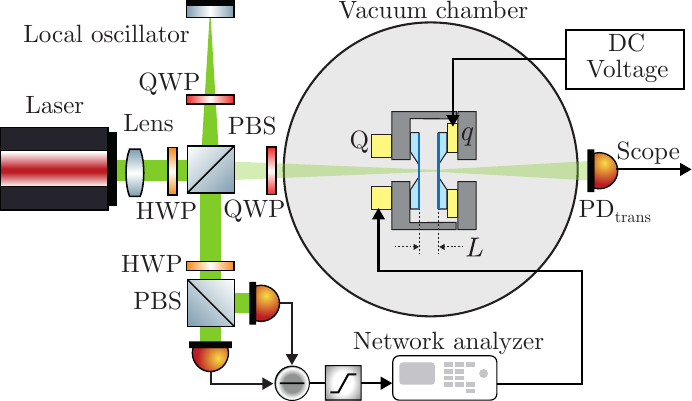}}
 \caption{Sketch of the experimental setup for characterizing the membrane--cavity. A \SI{532}{\nano\meter} laser is sent into a polarization--multiplexed Michelson interferometer. The field reflected by the two--membrane etalon contains information on the motion of the slabs, and it is detected by means of balanced homodyne detection. On the other side of the cavity, the transmitted light impinges directly on a photodiode. HWP denotes a half--waveplate, QWP a quarter--waveplate, and PBS a polarizing beam--splitter. Two piezoelectric elements, indicated by $q$ and $Q$, are attached to a single membrane and to the common support, respectively. $q$ allows to change the cavity length, $Q$ moves the center--of--mass.}
\label{fig:Figure_Experiment}
\end{center}
\end{figure}
\begin{figure*}[h]
\begin{center}
  {\includegraphics[scale=0.55]{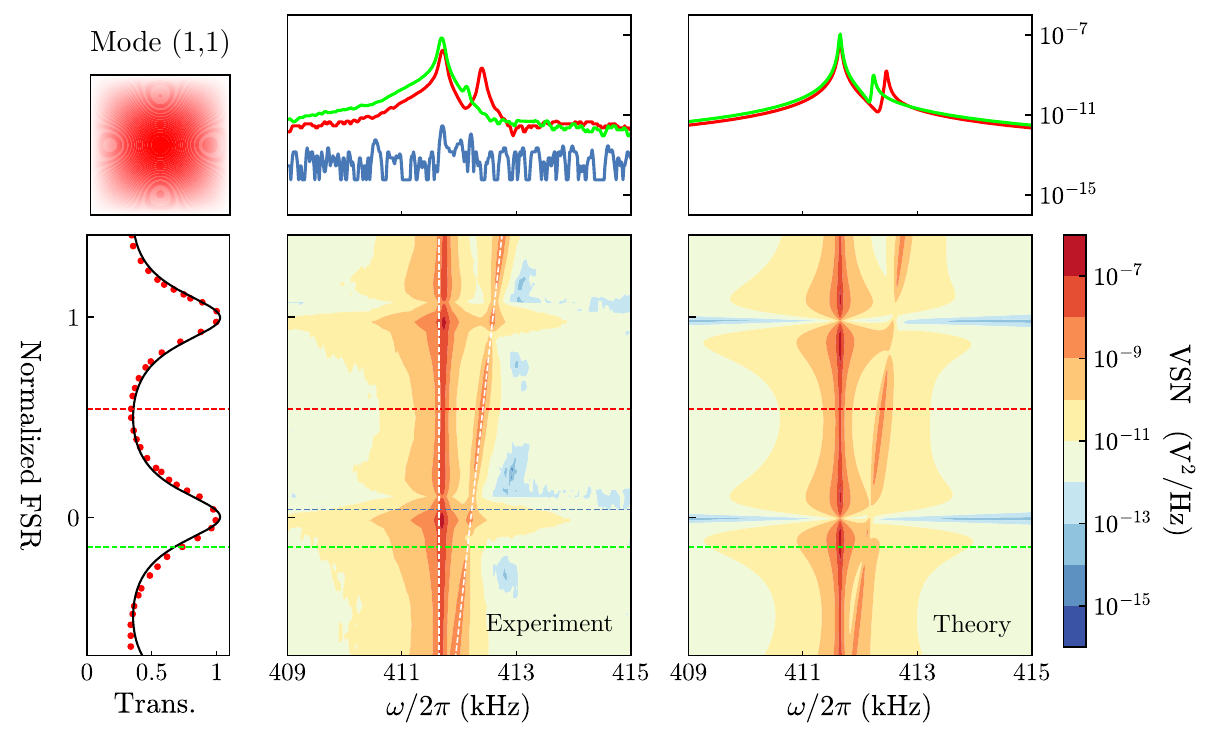}}
\caption{Homodyne spectra around the resonance of the mode (1,1) of the membranes. Different position of the $q$ piezo sweeps the free spectral range (FSR) of the etalon, represented here on the vertical axes, normalized for half the light wavelength. On the left panel the normalized transmitted intensity is shown. The measured data (red dots) are presented with Airy peaks (black line) with cavity finesse $\calF=2.809$. The middle and right panels show the measured and theoretical spectra, respectively. The white dashed lines on the experimental data mark the resonances. The green and red curves in the inset above represent the 2D spectra corresponding to the horizontal slice of the contour plot indicated with the same colors. The measured voltage spectral noise on the homodyne detector, without external excitation of the membranes, is presented in blue. In the simulation we use $\eta_1/\eta_2 = 0.1$.}
\label{fig:Mode11_complete}
\end{center}
\end{figure*}
\begin{figure*}[h]
\begin{center}
  {\includegraphics[scale=0.55]{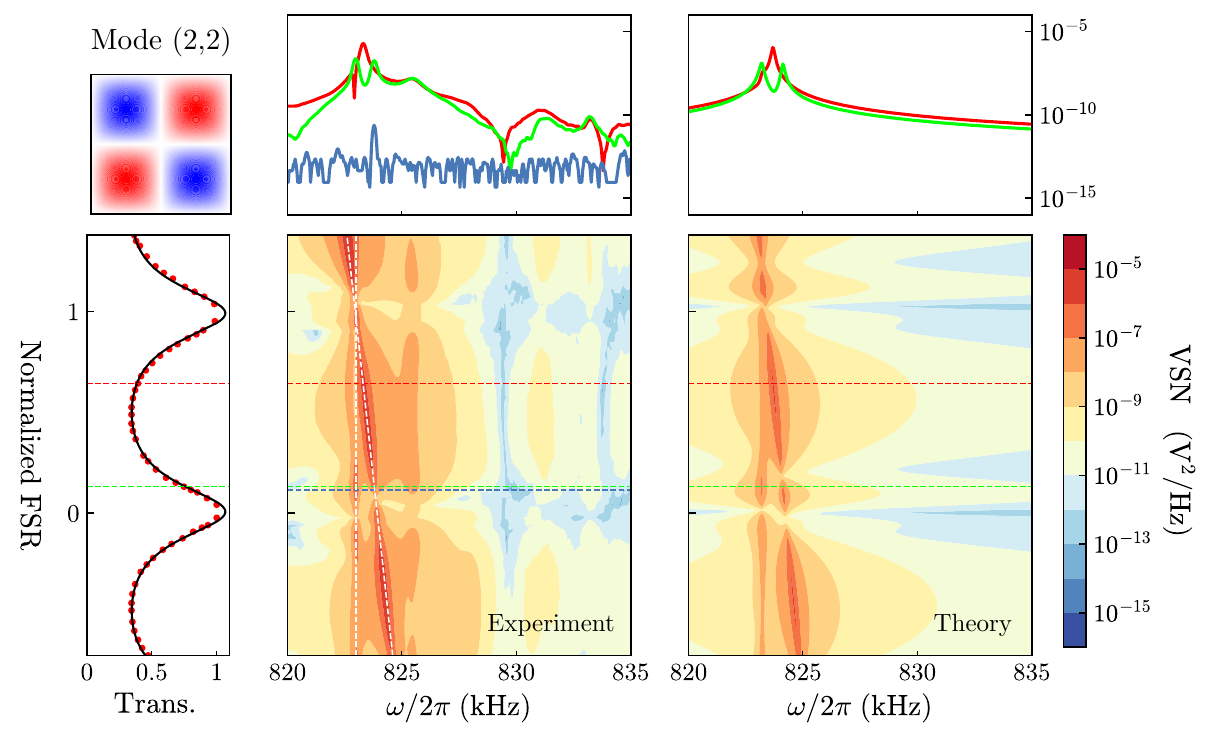}}
\caption{Homodyne spectra around the resonance of the mode (2,2) of the membranes. Different position of the $q$ piezo sweeps the free spectral range (FSR) of the etalon, represented here on the vertical axes, normalized for half the light wavelength. On the left panel the normalized transmitted intensity is shown. The measured data (red dots) are presented with Airy peaks (black line) with cavity finesse $\calF=2.809$. The middle and right panels show the measured and theoretical spectra, respectively. The white dashed lines on the experimental data mark the resonances. The green and red curves in the inset above represent the 2D spectra corresponding to the horizontal slice of the contour plot indicated with the same colors. The measured voltage spectral noise on the homodyne detector, without external excitation of the membranes, is presented in blue. In the simulation we use $\eta_1/\eta_2 = 6.7$.}
\label{fig:Mode22_complete}
\end{center}
\end{figure*}
The setup involved in the experiment is depicted schematically in Fig.~\ref{fig:Figure_Experiment}.
We have measured the spectrum of the field reflected by the cavity for different distances between the two membranes, i.e. different lengths of the cavity, which can be adjusted by applying a continuous voltage to the $q$ piezo.
The membranes are two, nominally identical, Si$_3$N$_4$ square membranes (the nominal side length is $\SI{1}{\milli\meter}$) by Norcada.
A complete characterization, described in details in the next section, reveals a side length of about $\SI{0.98}{\milli\meter}$, slab thickness $L_{\rm{m}}=\SI{75.2\pm0.3}{\nano\meter}$ and reflectivity $\calR=\num{0.3618\pm0.0003}$ at $\SI{532}{\nano\meter}$, for which the cavity has a finesse $\calF=\num{2.809\pm0.003}$.
The distance between the two membranes is estimated to be $L_{\rm{c}}=\SI{5.707\pm0.002}{\micro\meter}$.
The membrane--sandwich is mounted inside a vacuum chamber, evacuated at $\SI{4e-7}{\milli\bar}$, at environment temperature $\SI{293}{\kelvin}$.

The reflected field has been detected interferometrically, by means of balanced homodyne detection: we have adopted a network analyzer to shake the $Q$ piezo and measure the voltage spectral noise on the homodyne detector, which represents the closed--loop transfer function of the cavity to the piezo excitation.
On the other hand, the field transmitted by the cavity impinges directly on a photodiode, and it allows to reconstruct the typical cavity fringes, which are presented in Fig.~\ref{fig:Mode11_complete} and Fig.~\ref{fig:Mode22_complete} in the panels on the left.
The solid line represents the theoretical curve with finesse $\calF=2.809$.
Moreover, in Fig.~\ref{fig:Mode11_complete} and Fig.~\ref{fig:Mode22_complete}, we show the measured (middle panel), and expected contributions (right panel), according to Eq.~\eqref{eq:Sxx}, around the modes (1,1), and (2,2) of the mechanical oscillators, respectively.
The white dashed lines on the experimental data of Fig.~\ref{fig:Mode11_complete} and Fig.~\ref{fig:Mode22_complete} identify the resonances.
The frequency shifts as a function of the applied voltage, noticed in Fig.~\ref{fig:Mode11_complete} and Fig.~\ref{fig:Mode22_complete}, are ascribed to the bend of the piezo over which one membrane is glued; the piezo bends the frame varying the tensile stress on the membrane, which induces a variation of the mechanical eigenfrequencies \cite{PhysRevA.99.023851}.
We have considered this effect in the theoretical model as well, varying the resonance frequency $\omega_{mj}$ in the displacement spectrum, given by Eq.~\eqref{eq:displ_spectra}, of the mechanical resonator attached to the piezo.

Due to small misalignments, the mechanical modes are not equally illuminated.
The overlap between the optical field and the mechanical amplitude shape changes between different modes and membranes.
Comparing the simulations with the measurements, we can optimize the overlap parameters, $\eta_1$ and $\eta_2$, such that $\eta_1/\eta_2=0.1$ for the mode (1,1), and $\eta_1/\eta_2=6.7$ for the mode (2,2).
The external mechanical perturbation, coming from the $Q$ piezo, adds noise in the system, whose intensity is much larger than the thermal fluctuations. 
It can be controlled through the amplitude of the network analyzer source, which has been fixed at $-50\,\rm{dBm}$ for all the experimental runs.
The simulations, obtained from Eq.~\eqref{eq:Sxx}, are matched to the levels measured in the spectra taking into account the response of the piezo, and the transfer function of the system to the external vibration.
We observe noise cancellations in the spectra in Fig.~\ref{fig:Mode11_complete} and Fig.~\ref{fig:Mode22_complete}.
They emerge from the interference between higher order mechanical modes, which are not under investigation in this work.

Since the back--action of the light onto the mechanics is negligible in the experiment, the motion of the mechanical modes is not affected by the position of the membranes. 
On the contrary, the field reflected by the etalon and the amplitude of the homodyne fringe depend on the motion of the membranes and on the length of the cavity.
For this reason, even if the amplitude of the mechanical motion does not change, the distance within the two membranes influences the measured spectra.
The simulations point out accurately this behavior.
\section{Two--membrane etalon characterization}
\subsection{Mechanical characterization}
\begin{figure}[htbp]
\includegraphics[width=0.8\linewidth]{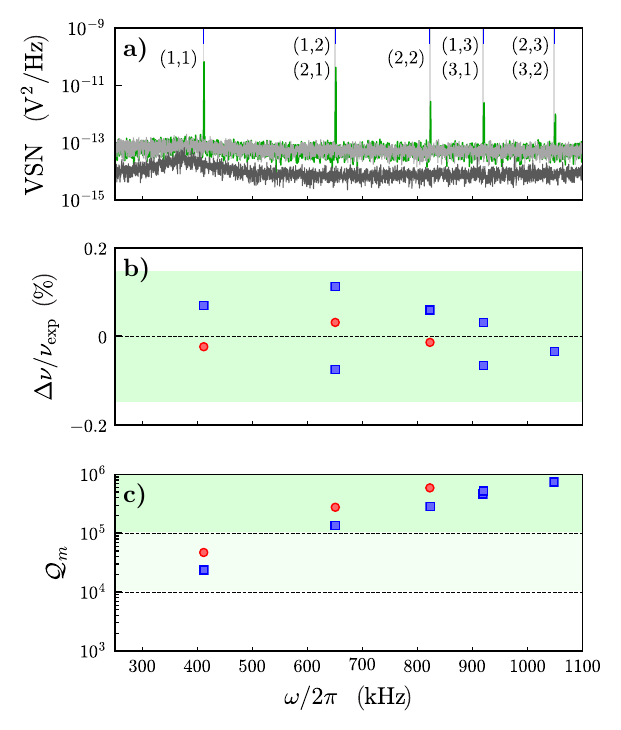}
\caption{Mechanical characterization of the membrane--sandwich. a) Homodyne voltage spectral noise (VSN), up to $\SI{1.1}{\mega\hertz}$. The detected mechanical peaks, referred to the modes $\mathrm{(n,m)}$ of the two membranes, are green, with the shot noise in grey, and the electronic noise floor in black. The blue ticks on the top of the figure indicate the theoretical normal modes frequency for a square membrane, with side length $\SI{1}{\milli\meter}$, stress and density $\sigma=\SI{1}{\giga\pascal}$, and $\rho=\SI{3100}{\kilo\gram\per\cubic\meter}$, respectively. b) Relative shift between the measured and the theoretical frequencies. Blue squares and red circles refer to the upper frequency and lower frequency membrane, respectively. c) Estimated mechanical quality factor $Q_m$ for the first normal modes of the oscillators.}
\label{fig:mech_car}
\end{figure}

In this section we carry out the characterization of the mechanical oscillators.
In order to determine the resonance frequencies of the first normal modes of vibration of the two membranes, we measure the thermal voltage spectral noise on the homodyne detector, i.e. the uncorrelated noise on the detector when the piezo $Q$ is shorted.
In Fig.~\ref{fig:mech_car}a) one can observe the measured data, where the green peaks are the normal modes of the membranes.
The resonance frequency of the vibrational normal mode $(n,m)$ of a square membrane can be calculated as
\begin{equation}
\nu_{(n,m)}=\sqrt{\frac{\sigma}{\rho}\left[\left(\frac{n}{L_{\rm{x}}}\right)^2+\left(\frac{m}{L_{\rm{y}}}\right)^2\right]},
\label{eq:norm_modes}
\end{equation}
where $L_{\rm{x}}$ and $L_{\rm{y}}$ are the side lengths of the membrane, and we assume the nominal values for the tensile stress, and for the density, $\sigma=\SI{1}{\giga\pascal}$, and $\rho=\SI{3100}{\kilo\gram\per\cubic\meter}$, respectively.
From a comparison between the expected frequencies and the measured values, with their relative shift shown in Fig.~\ref{fig:mech_car}b), we determine the sizes $L_{\rm{x}}^{(1)}=\SI{0.9774\pm0.0001}{\milli\meter}$, $L_{\rm{y}}^{(1)}=\SI{0.9759\pm0.0001}{\milli\meter}$, for one membrane, and $L_{\rm{x}}^{(2)}=\SI{0.9756\pm0.0001}{\milli\meter}$, $L_{\rm{y}}^{(2)}=\SI{0.9773\pm0.0001}{\milli\meter}$, for the other.
By fitting the mechanical peaks with a Lorentzian function we can extract the damping rates.
The mechanical quality factor, in the limit of good--oscillator, that is for resonance frequency much larger than the bandwidth, can be obtained from the ratio between its resonance frequency and its damping rate.
The estimated quality factors for the first measured modes are collected in Fig.~\ref{fig:mech_car}c).

\subsection{Optical properties}
\begin{figure}[htbp]
\includegraphics[width=0.8\linewidth]{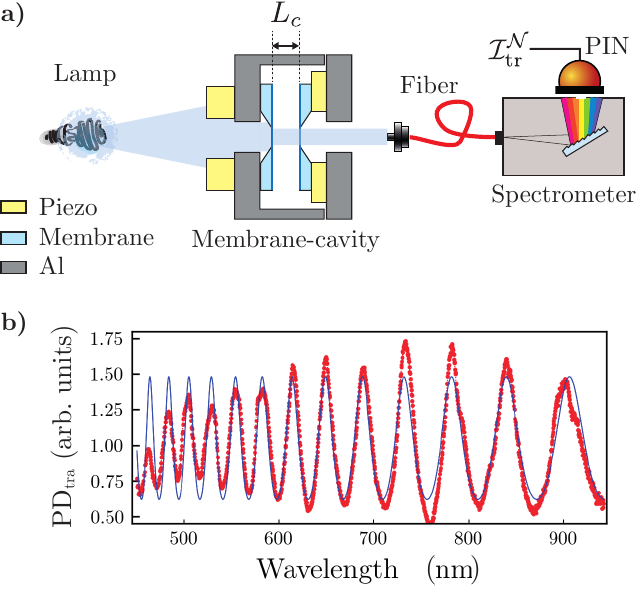}
\caption{Cavity frequency scan. a) The light emitted by a tungsten bulb illuminates the cavity. The transmitted light is collected in a multi--mode optical fiber, and frequency analysed by means of a spectrometer. b) The measured transmitted intensity as a function of the wavelength is shown as red points, with the best--fitting Airy peaks as a blue line. The estimated cavity length is $L_{\rm{c}}=\SI{5.707\pm0.002}{\micro\meter}$.}
\label{fig:cav_whitelamp}
\end{figure}

\begin{figure}[htbp]
\includegraphics[width=\linewidth]{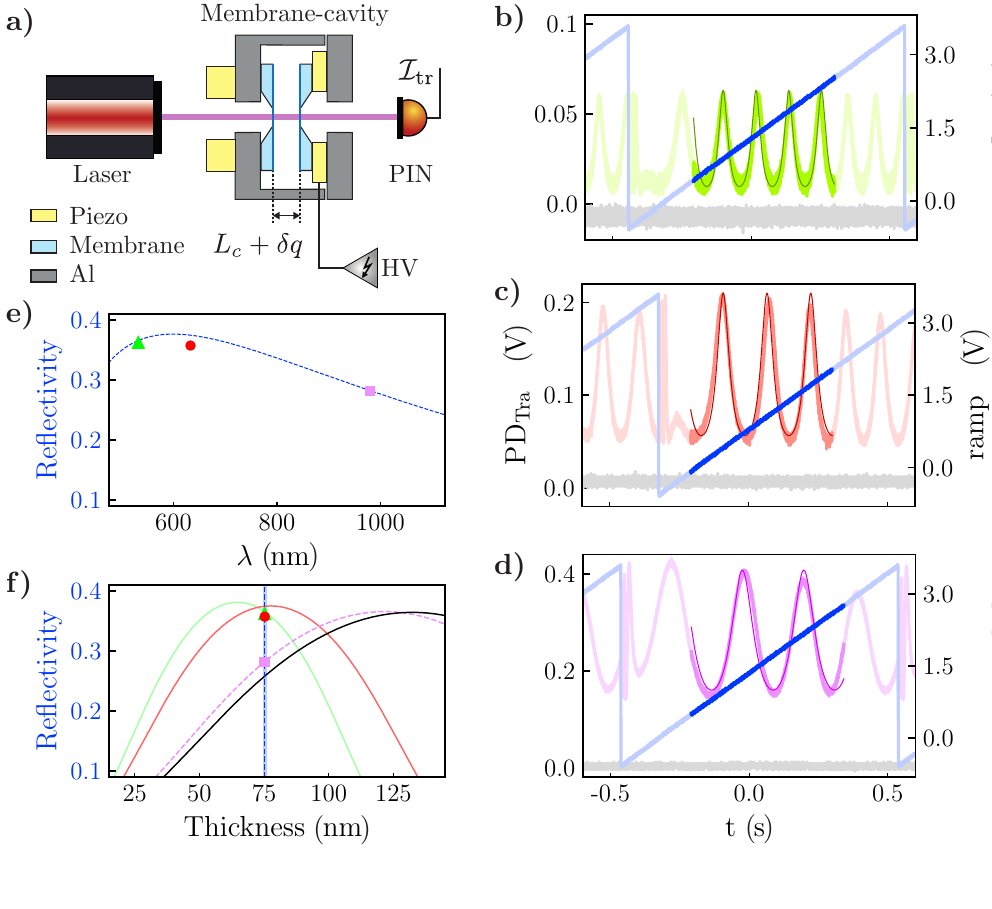}
\caption{Cavity time scan. a) The cavity length is swept applying a high--voltage (HV) ramp to the $q$ piezo. A PIN photodiode directly detects the transmitted light. The experiment is repeated for three different laser wavelength, $\SI{532}{\nano\meter}$ b), $\SI{632.8}{\nano\meter}$ c), and $\SI{980}{\nano\meter}$ d). The voltage ramp is shown as blue points, together with the measured data (colored points), and the best--fitting curve (solid line). From the analysis we find the finesse coefficients $\calF_{532}=\num{2.809\pm0.003}$, $\calF_{632.8}=\num{2.766\pm0.003}$, and $\calF_{980}=\num{1.9774\pm0.0008}$. Assuming identical reflectivities for the two membranes, we calculate $\calR_{532}=\num{0.3618\pm0.0003}$, $\calR_{632.8}=\num{0.3571\pm0.0003}$, and $\calR_{980}=\num{0.2652\pm0.0001}$. e) The dependence of the Si$_3$N$_4$ slabs reflectivity on the light wavelength is shown, with the green triangle, the red circle, and the purple square indicating the measured data at $\SI{532}{\nano\meter}$, $\SI{632.8}{\nano\meter}$, and $\SI{980}{\nano\meter}$, respectively. The blue dashed line is the best--fitting function from Eq.~\eqref{eq:rm}, giving the slab thickness $L_{\rm{m}}=\SI{75.2\pm0.3}{\nano\meter}$. f) The dependence of the reflectivity on the thickness of the slabs is represented, using the same symbol as before for the experimental points, the dashed blue line indicates the best--fitted value, and the lines show the behavior for different laser wavelengths. The green, the red, and the purple ones refer to $\SI{532}{\nano\meter}$, $\SI{632.8}{\nano\meter}$, and $\SI{980}{\nano\meter}$, respectively. The black one stands for $\SI{1064}{\nano\meter}$.}
\label{fig:memb_refl}
\end{figure}

The length of the membrane--cavity, its finesse coefficient, and the width of the membranes are investigated here.
As shown in Fig.~\ref{fig:cav_whitelamp}a), exploiting the white light of a tungsten lamp, which illuminates the cavity, we can measure the transmitted light spectrum over a wide range of wavelength.
Fitting the result with Airy peaks, see Fig.~\ref{fig:cav_whitelamp}b), we determine the distance within the two membranes $L_{\rm{c}}=\SI{5.707\pm0.002}{\micro\meter}$.
On the other hand, for studying the reflectivity of the sandwich, we perform a time scan of the cavity, sweeping its length, at different light wavelength ($\SI{532}{\nano\meter}$, $\SI{632.8}{\nano\meter}$, and $\SI{980}{\nano\meter}$), as described in Fig.~\ref{fig:memb_refl}. 
Fitting the peaks in the transmission signal we find the finesse coefficients for the three wavelengths $\calF_{532}=\num{2.809\pm0.003}$, $\calF_{632.8}=\num{2.766\pm0.003}$, and $\calF_{980}=\num{1.9774\pm0.0008}$, respectively, which are associated to the reflectivity values $\calR_{532}=\num{0.3618\pm0.0003}$, $\calR_{632.8}=\num{0.3571\pm0.0003}$, and $\calR_{980}=\num{0.2652\pm0.0001}$.
We assume the two membranes have identical reflectivities.
Eventually, knowing the reflectivity of the dielectric slab for different wavelengths, we can estimate the thickness $L_{\rm{m}}=\SI{75.2\pm0.3}{\nano\meter}$.
\section{Conclusions}
In this work we have analyzed the fields of a two--membrane etalon device, driven by a laser.
We detect interferometrically the reflected field, and we measure the intensity of the transmitted field, for different distances within the two slabs. 
The measured spectra present modulations of the intensity of the mechanical peaks, which depends on the length of the cavity.
The theoretical model, developed in the paper, provides simulations in good agreement with the experimental results.
We have also presented a general theory of the input, intra--cavity and output fields, effective for any value of the cavity finesse and which includes the effect of the membrane motion.
It reproduces all the known results in the literature valid in the limit of high--finesse and/or without the membrane motion.
The ductility of the model allows to treat a general two slabs etalon, which is a powerful tool for optomechanical applications where it is coupled to a high--finesse Fabry-P\'{e}rot cavity.
This optomechanical setup allows many achievements, and the theoretical model proposed in this work is promising to describe it analytically.

\section{Acknowledgements}
We acknowledge the support of the PNRR MUR project PE0000023-NQSTI (Italy).


\begin{thebibliography}{9}

\bibitem{Kogelnik:66}
	H.~Kogelnik and T.~Li,
	Appl.~Opt.,
	{\bf 10},
	(5) 
	1966.

\bibitem{Siegman:1999}
	A.~E.~Siegman,
	IEEE~Journal~of~Selected~Topics~in~Quantum~Electronics,
	{\bf 6},
	(6) 
	2000.
	
\bibitem{YE20031}
	J.~Ye and T.~W.~Lynn,
	\newblock {\em Applications of Optical Cavities in Modern Atomic, Molecular, and Optical Physics}.
	\newblock Academic Press, 2003.
	
\bibitem{PhysRevA.12.148}
	k.~Ujihara,
	Phys.~Rev.~A,
	{\bf 12},
	(148--158) 
	1975.

\bibitem{Kim:1994vs}
	M.~Kim et.~\underline{al.},
	Phys.~Rev.~A,
	{\bf 50},
	(5) 
	1994.
	
\bibitem{Aiello:2000wn}
	A.~Aiello,
	Phys.~Rev.~A,
	{\bf 62},
	(6) 
	2000.
	
\bibitem{Viviescas:2003}
	C.~Viviescas and G.~Hackenbroich,
	Phys.~Rev.~A,
	{\bf 67},
	(013805) 
	2003.
	
\bibitem{Aspelmeyer:2014vr}
	M.~Aspelmeyer et.~\underline{al.}, 
	Rev.~Mod.~Phys.,
	{\bf 86},
	(1391) 
	2014.

\bibitem{Teufel:2011vw}
	J.~D.~Teufel et.~\underline{al.}, 
	Nature,
	{\bf 475},
	(359--363) 
	2011.
	
\bibitem{Verhagen:2012aa}
	E.~Verhagen et.~\underline{al.}, 
	Nature,
	{\bf 482},
	(63--67) 
	2012.
	
\bibitem{Brubaker:2022aa}
	B.~M.~Brubaker et.~\underline{al.}, 
	Phys.~Rev.~X,
	{\bf 12},
	(021062) 
	2022.
	
\bibitem{Rossi:2018ve}
	M.~Rossi et.~\underline{al.}, 
	Nature,
	{\bf 563},
	(53--58) 
	2018.

\bibitem{Mason:2019aa}
	D.~Mason et.~\underline{al.}, 
	Nat.~Phys.,
	{\bf 15},
	(745--749) 
	2019.
	
\bibitem{Barzanjeh:2022aa}
	S.~Barzanjeh et.~\underline{al.}, 
	Nat.~Phys.,
	{\bf 18},
	(15--24) 
	2022.
	
\bibitem{Thompson:2008aa}
	J.~D.~Thompson et.~\underline{al.}, 
	Nature,
	{\bf 452},
	(72--75) 
	2008.
	
\bibitem{Jayich_2008}
	A.~M.~Jayich et.~\underline{al.}, 
	New~J.~Phys.,
	{\bf 10},
	(095008) 
	2008.
	
\bibitem{PhysRevLett.103.207204}
	D.~J.~Wilson et.~\underline{al.}, 
	Phys.~Rev.~Lett.,
	{\bf 103},
	(207204) 
	2009.
	
\bibitem{PhysRevA.78.041801}
	M.~Bhattacharya and P.~Meystre, 
	Phys.~Rev.~A,
	{\bf 78},
	(041801) 
	2008.
	
\bibitem{Gartner:2018aa}
	C.~Gärtner et.~\underline{al.}, 
	Nano Letters,
	{\bf 18},
	(7171--7175) 
	2018.
	
\bibitem{PhysRevA.99.023851}
	X.~Wei et.~\underline{al.}, 
	Phys. Rev. A,
	{\bf 99},
	(023851) 
	2019.

\bibitem{Piergentili:2018aa}
	P.~Piergentili et.~\underline{al.}, 
	New~J.~Phys.,
	{\bf 18},
	(083024) 
	2018.
	
\bibitem{PhysRevX.15.011014}
	X.~Yao et.~\underline{al.}, 
	Phys. Rev. X,
	{\bf 15},
	(011014)
	2025.

\bibitem{Yang:2020aa}
	C.~Yang et.~\underline{al.}, 
	Nat.~Commun.,
	{\bf 11},
	(4656) 
	2020.
	
\bibitem{Piergentili:2021}
	P.~Piergentili et.~\underline{al.}, 
	New~J.~Phys.,
	{\bf 23},
	(073013) 
	2021.
	
\bibitem{PhysRevApplied.15.034012}
	P.~Piergentili et.~\underline{al.}, 
	Phys. Rev. Applied,
	{\bf 15},
	(034012) 
	2021.
	
\bibitem{Piergentili:2022aa}
	P.~Piergentili et.~\underline{al.}, 
	Photonics,
	{\bf 9},
	(99) 
	2022.
	
\bibitem{10.3389/fphy.2023.1222056}
	F.~Marzioni et.~\underline{al.}, 
	Front.~Phys.,
	{\bf 11},
	2023
	
\bibitem{Xu:2016aa}
	H.~Xu et.~\underline{al.}, 
	Nature,
	{\bf 537},
	(80--83) 
	2016.

\bibitem{Schliesser:2008hc}
	A.~Schliesser et.~\underline{al.}, 
	Nat.~Phys.,
	{\bf 4},
	(415) 2008.

\bibitem{Reinhardt:2017ab}
	C.~Reinhardt et.~\underline{al.}, 
	Opt.~Express,
	{\bf 25},
	(1582) 2017.

\bibitem{Saleh:2007sw}
	B.~E.~A.~Saleh and M.~C.~Teich,
	\newblock {\em Fundamentals of Photonics, 2nd edition}.
	\newblock Wiley, New York, 2007.

\bibitem{Sacher:2008aa}
	W.~D.~Sacher and J.~K.~S.~Poon,
	Opt.~Express,
	 {\bf 16}
	 (15741--15753) 
	 2008.

\bibitem{Crosignani:1986aa}
	B.~Crosignani, A.~Yariv, and P.~Di~Porto,
	Opt.~Lett.,
	{\bf 11},
	(251--253) 
	1986.

\bibitem{Krasnov1983:aa}
	M.~L.~Krasnov, A.~I.~Kiselev, and G~I.~Makarenko,
	\newblock {\em Integral equations, 1st edition}.
	\newblock Mir, Moskov 1983, 2007.

\bibitem{Tsubono:1991aa}
	K.~Tsubono, N.~Mio, and A.~Mizutani,
	Japan.~J.~App.~Phys., 
	{\bf 6}
	(1326--1330)
	1991.

\bibitem{Raymer:2013aa}
	M.~G.~Raymer and C.~J.~McKinstrie,
	Phys.~Rev.~A, 
	{\bf 88}
	(043819) 
	2013.

\end{thebibliography}
\end{document}